\documentclass[12pt]{article}
\usepackage{amsmath}
\usepackage{amssymb}
\newcommand{\Master}{Flow-Along Equation}
\newcommand{\W}{\Omega}
\newcommand{\w}{W}
\newcommand{\Proj}[1]{\mathcal{P}_{#1}}

\newcommand{\Lie}[2]{\underset{#1}{\mathcal{L}} {#2}}
\newcommand{\ti}{\Gamma}
\renewcommand{\SS}{\scriptscriptstyle}
\newcommand{\eexp}[1]{{\rm e}^{#1}}
\newcommand{\opeq}{\succ}
\title{Lagrangian, Hamiltonian and other Structures for the Heat Equation and Potential Burgers Equation}
\author{Miguel D Bustamante${}^{1}$ and Sergio A Hojman${}^{1,\, 2,\, 3}$\\
${}^1$ Departamento de F\'{\i}sica, Facultad de Ciencias, Universidad de Chile,\\
Santiago, Chile\\ ${}^2$ Facultad de Educaci\'on, Universidad Nacional Andr\'es
Bello,\\ Santiago, Chile\\ ${}^3$ Centro de Recursos Educativos Avanzados, CREA,\\
Santiago, Chile \\  E-mail:\,{\bf
miguelb@macul.ciencias.uchile.cl\,,\,\,shojman@ctcreuna.cl}}
\date{September 25, 2001}
\begin{document}
\maketitle
\begin{abstract}
 In this work, we construct the general solution to the Heat Equation (HE)
and to many tensor structures associated to the Heat Equation, such as Symmetries,
Lagrangians, Poisson Brackets (PB) and Lagrange Brackets (LB), using newly devised
techniques that may be applied to any linear equation (e.g., Schr\"odinger Equation in
field theory, or the small-oscillations problem in mechanics). In particular, we
improve (increase the rank of) a time-independent PB found recently \cite{mauricio}
which defines a Hamiltonian Structure for the HE, and we construct an Action Principle
for the HE. We also find a new structure, which we call a Metric Structure (MS), which
may be used to define alternative anti-commutative ``Hamiltonian" theories, in which
the Metric- or M-Hamiltonians have to be explicitly time-dependent. Finally, we map
some of these results to the Potential Burgers Equation (PBE).
\end{abstract}
\newpage
\section{Introduction} The Heat Equation (HE) $\omega_t(x,t) = \omega_{xx}(x,t)$ is
an instructive case of an integrable equation, for it is easier to find Lagrangian
Structures \cite{hojmanLagr} than Hamiltonian Structures \cite{hojman1} for that
equation. In this paper we construct several tensor dynamical structures related to it
including a symmetric Metric Structure operator which allows for an anti-commutative
Hamiltonian-like evolution representation of the system. An improved version of a
Hamiltonian System for the HE \cite{mauricio}, with a seemingly regular Poisson Bracket
is found. Furthermore, several regular Action Principles for the HE are presented.
Finally, we map the obtained results into the Potential Burgers Equation (PBE).

In section \ref{secPrev} we summarize useful known results as well as some new ones. In
section \ref{secLagr} we start by developing the Lagrangian approach \cite{hojmanLagr}
to the HE, which is easier to deal with than the Hamiltonian one in the sense that
there is no problem with time-dependent Lagrangians to define Action Principles. We
will see that Action Principles may be found such that their Euler-Lagrange equations
are equivalent to the HE; we think this is the first Lagrangian formalism ever found
for the Heat Equation. The Lagrange Brackets (LB) for these Action Principles are
explicitly time-dependent, thus there is no Hamiltonian Structure related to them
\cite{hojman2}.

From the contraction\footnote{A contraction of two tensors is a sum$\otimes$integral
over repeated indices: for example, the contraction between the $(2,0)$ tensor $J$ and
the 1-form $U$ is  the vector with components $\eta^x = J^{x y} U_y$.}
 between the LB of the Lagrangian approach just mentioned and a
known Strong Symmetry for the HE \cite{mauricio}, we find a new structure, consisting
of a symmetric $(2,0)$ tensor which we call a Metric, and a time-dependent Constant of
the Motion (Metric- or M-Hamiltonian); this Metric Structure (MS) defines an
anti-commutative formalism. This will be performed in section \ref{secMS}.

Next, in section \ref{secGeSo} we make use of the seemingly trivial general solution to
the HE, to build systematically non-trivial Symmetries, Lagrangians, LBs, PBs, MSs and
Strong Symmetries, in a way that generalizes the above constructions.

An interesting method to find Hamiltonian Structures is applied recently in
\cite{mauricio} for the HE, nevertheless even though the ranks of the PBs found are
infinite, they are not maximal. The method needs a set of time-independent Symmetries
of the HE such that the Lie Derivative \cite{Nakahara} of a Constant of the Motion (the
Hamiltonian) for the HE along any Symmetry in that set is zero \cite{hojman1}. In
section \ref{secPB}, a larger set is found using the general solution, and thus we
increase the rank of the PB, which now seems to be maximal.

Finally, using the Cole-Hopf coordinate transformation, we map the HE and some of its
related structures into the Potential Burgers Equation (PBE), thus
 we show new LB, PB, MS, Symmetries and Lagrangian Structures for the PBE
in section \ref{secPBE}.

We remark that all these results  may be mapped directly into the Burgers Equation, and
that the techniques used for the Heat Equation may be performed on any linear evolution
equation, e.g., the Schr\"odinger Equation and the finite-dimensional
small-oscillations problem.
\newpage

\section{Preview and Notation} \label{secPrev}
\subsection{Special Symbols: `` $\doteq$'' and `` $\opeq$''}

Throughout this paper, tensor structures in infinite dimensional spaces are presented
in specific coordinate systems. Taking the risk of being too sharp, we intend to
clarify the different representations with an example taken from Quantum Mechanics:
consider the $x$-momentum, ${\bf \hat{p}}$. This operator is considered as an invariant
structure, which acts upon Wavefunctions linearly. In our formalism, Wavefunction ${\bf
\w}$ is a vector and typical operators are $(1,1)$ tensors, all of which are invariant
structures on $\mathbb{C}$. When we represent these structures in a specific coordinate
system, say the ``$x$-representation" or $\psi$-coordinate system, we write ($
\hbar=1$):

\begin{equation}\label{qm}
\begin{array}{rrcl}
 {\rm Coordinate\,\, system:\,}& \W^{\psi(x)} & \doteq & \psi(x,t) \\
 {\rm Wavefunction\, vector:\,}& \w^{\psi(x)} & \doteq & \psi(x,t) \\
 { \rm Momentum\,} (1,1){\rm \,tensor:\,}&   \hat{p}_{\quad\psi(y)}^{\psi(x)} & \doteq &- i \,\delta'(x-y)\,.
\end{array}
\end{equation}
Notice that, in the LHS, $\psi(x)$ is just a label, a discrete$\otimes$continuous index
in the vector space in which \mbox{${\bf \w} \equiv \w^{\psi(x)}\frac{\delta}{\delta
\W^{\psi(x)}}$} is defined (Einstein summation convention over repeated indices is
assumed). In the $\psi$-coordinate system, the vector ${\bf \hat{p}\cdot \w}$ (which
defines the momentum of the field) is written as a contraction:
\[
 \hat{p}_{\quad\psi(y)}^{\psi(x)} \w^{\psi(y)}  \doteq  -{\rm i}  \int dy\, \delta'(x-y)\psi(y,t) = -{\rm i}  \,\frac{\partial \psi}{\partial x}(x,t)\,,
\]
which means that the $(1,1)$ tensor ${\bf \hat{p}}$ has the following local expression
as an operator:

\begin{equation}\label{qm2}
 \begin{array}{rrcl}
{\rm \quad Operator \,expression:\, } & \hat{p}[\psi]& \opeq& -{\rm i} \, D\, ,
 \end{array}
 \end{equation}
where $D \equiv \frac{\partial}{\partial x}$ is the ``$x$-derivative operator''
\cite{derivOp}.

Equations of type (\ref{qm}) and (\ref{qm2}) are thus local (i.e.,
coordinate-dependent) representations of the invariant structures. In order to see the
above structures in the ``\mbox{$p$-representation}" or $\phi$-coordinate system
instead, we have to map with the transformation matrix $F_{\quad\psi(x)}^{\phi(p)} =
\frac{\delta \W^{\phi(p)}}{\,\,\delta \W^{\psi(x)}}\equiv \frac{ \exp( {\rm i} p
x)}{\sqrt{\rm 2 \pi }}$ (the Fourier transform). We get:
\[
\begin{array}{rrcl}
  {\rm  Coordinate \,\, system:\,}&\W^{\phi(p)} &\doteq & \phi(p,t) \equiv \int
dx\,\frac{ \exp( {\rm i} p x)}{\sqrt{\rm 2 \pi }} \psi(x,t) \\
  {\rm  Wavefunction\, vector:\,}&\w^{\phi(p)} &\doteq & \phi(p,t) \\
                                    &\left(\w_{}^{\phi(p)}\right.&\,\,\equiv & F_{\quad\psi(x)}^{\phi(p)}\left. \w^{\psi(x)} \right)  \\
 { \rm Momentum\,} (1,1){\rm \,tensor:\,}&  \hat{p}_{\quad\phi(q)}^{\phi(p)} &\doteq & p \,\delta(p-q)\\
                                    &\left(\hat{p}_{\quad\phi(q)}^{\phi(p)}\right.&\,\,\equiv &
                                    \left. F_{\quad\psi(x)}^{\phi(p)} \hat{p}_{\quad\psi(y)}^{\psi(x)}
                                    (F^{\SS -1})_{\quad\phi(q)}^{\psi(y)}
                                    \right)\\
{\rm Operator \,expression:\,}&  \hat{p}[\phi]      &\opeq  & p\,\mathbb{I}\\
                                &\left(\hat{p}[\phi] \right. &\,\,\equiv&\left. F
\hat{p}[\psi]F^{\SS -1}\right) \,,
\end{array}
\]
 where $\mathbb{I}$ is the Identity operator. The momentum ${\bf \hat{p}\cdot \w}$
 is written in the $\phi$-coordinate system:
 \[
{\bf \hat{p}\cdot \w}^{\phi(p)} \equiv \hat{p}_{\quad\phi(q)}^{\phi(p)}\w^{\phi(q)}
\doteq p \,\phi(p,t)\,.
 \]  We remark that an operator expression
may be defined for type $(2,0)$ and $(0,2)$ tensors as well. This will be the case of
Poisson and Lagrange Brackets, and Metrics.

\subsection{The Lagrangian Structure}

We are going to discuss the construction of Action Principles for field equations. Take
for example the Heat Equation:\[\frac{\partial \omega}{\partial t }(x,t)
=\frac{\partial^2 \omega}{\partial x^2}(x,t).\] This is a first order (in time)
equation for one real field ($\omega$), and one spatial dimension ($x$). Let us promote
the $x$-coordinate to a continuous index, by means of the representation:
\[
\W^{\omega(x)} \doteq \omega(x,t)\, .
\]
This is merely the definition of the coordinates $\W^{\omega(x)}$ for every $x$. Let us
define the vector ${\bf V}$ by its components in the $\omega$-coordinate system:
\[
 V^{\omega(x)} \doteq \frac{\partial^2 \omega}{\partial x^2}(x,t)\,.
 \]
We are dealing with an infinite-dimensional vector space over $\mathbb{R}$. The Heat
Equation, therefore, can be cast into the general form of an autonomous system of
first-order \emph{equations of motion} for one real field,

\begin{equation}
\label{motion}
 \frac{{\rm d} \W^{ \psi(\mathbf{x})}}{{\rm d} t} = V^{ \psi(\mathbf{x})}\big[\W^{
\psi(\mathbf{y})}\big] \quad ,
\end{equation}
where $\psi(\mathbf{x})$ is a multi-label consisting of:  a \emph{discrete} ($\psi$)
label, representing the coordinate system, and a set of \emph{continuous}
($\mathbf{x}=(x_1,x_2,\ldots)$) indices (the set of independent variables excluding
time); $V^{\psi(\mathbf{x})}$ are the components of the vector ${\bf
V}=V^{\psi(\mathbf{x})} \frac{\delta}{\delta \W^{\psi(\mathbf{x})}}$ in the
$\psi$-coordinate system. The Heat Equation is obtained setting the label $\psi =
\omega$, i.e., in the \mbox{$\omega$-coordinate} system, while the Potential Burguers
Equation is obtained setting $\psi = u$, where $\omega(x,t) = \eexp{u(x,t)}$ is the
change of coordinates (see section \ref{secPBE}). We remark that this formalism may be
extended straightforwardly to the case of many interacting fields.

In this setting, it can be shown \cite{hojmanLagr} that the equations of motion
(\ref{motion}) are related to a Variational Principle, given by

\begin{equation} \label{action}
 \mathcal{S} =
\int_{t_0}^{t_1} {\rm d} t\,L_{ \psi(\mathbf{x})}\big[\W^{
\psi(\mathbf{y})}(t),t\big]~\left( \frac{{\rm d} \W^{ \psi(\mathbf{x})}}{{\rm d} t} -
V^{ \psi(\mathbf{x})}\big[\W^{ \psi(\mathbf{y})}(t)\big] \right),
\end{equation}
where the integration over repeated continuous indices is assumed. Here,
$L_{\psi(\mathbf{x})}\big[\W^{\psi(\mathbf{y})}(t),t\big]$ are the components of the
Lagrangian 1-form, which obeys the following equation:

\begin{equation}
\label{lagrangian}
 \frac{\partial L_{\psi(\mathbf{x})}}{\partial t} +  L_{\psi(\mathbf{x}),\psi(\mathbf{y})}
  V^{\psi(\mathbf{y})} + L_{\psi(\mathbf{y})} V^{\psi(\mathbf{y})}{}_{, \psi(\mathbf{x})}  =
 0~,
\end{equation}
or
 \[\begin{array}{rcl}
 \bigg( \frac{\partial}{\partial t} + \Lie{\mathbf{V}}{} \bigg) \mathbf{L}& = &0 \\
\mathbf{L}& \equiv & L_{\psi(\mathbf{x})} \delta \W^{\psi(\mathbf{x})}\, ,
 \end{array}
 \] where $L_{\psi(\mathbf{x}),
\psi(\mathbf{y})} \equiv \frac{\delta L_{\psi(\mathbf{x})}}{\delta
\W^{\psi(\mathbf{y})}} $ is the functional derivative, $ \frac{\partial}{\partial t}$
denotes a derivative on the explicit dependence on time (i.e., under constant
$\W^{\psi(\mathbf{x})}$) and $\Lie{\mathbf{V}}{}$ is the Lie derivative \cite{Nakahara}
along the vector $\mathbf{V}$.

  The Euler-Lagrange equations which come from the action (\ref{action}) are:
\[
 \Sigma_{\psi(\mathbf{x}) \psi(\mathbf{y})}\left( \frac{{\rm d}\W^{\psi(\mathbf{y})}}{{\rm d}
t} - V^{\psi(\mathbf{y})} \right) = 0 , \] where $\Sigma_{\psi(\mathbf{x})
\psi(\mathbf{y})}\equiv L_{\psi(\mathbf{y}) , \psi(\mathbf{x})} - L_{\psi(\mathbf{x}),
\psi(\mathbf{y})}$ are the components of the 2-form Lagrange Bracket (LB)
$\mathbf{\Sigma} \equiv \delta \mathbf{L}$; it is closed under exterior derivation
\cite{Nakahara} ($\delta \mathbf{\Sigma} \equiv 0$) by definition and it obeys the
\Master\footnote{We use the name ``\Master" here for the first time, though this
equation has been already introduced in the respective references.}
\cite{hojmanLagr,hojman2}:
\[
\bigg( \frac{\partial}{\partial t} + \Lie{\mathbf{V}}{} \bigg)\mathbf{\Sigma} = 0 \, .
\]
These structures are not unique, for there may be different Lagrangians (not differing
by a total derivative) for the same equation of motion \cite{Shepley}.  We construct
some of them using the Projector, which is defined in the next section. It should be
noted that the LB must have no Kernel in order that the Action Principle (\ref{action})
be equivalent to the equations of motion (\ref{motion}), otherwise we would get
``deformed" equations, the deformation being related to the Kernel of the LB (this is
an interesting issue, anyway).
\subsection{The Projector}

For \emph{any\,} tensorial object we may associate its \mbox{\bf Projection} to a
solution of the \Master  \hspace{0truemm} for some vector $\mathbf{V}$.
 We define the Projector $\Proj{\mathbf{V}}$ as:
\[
\Proj{\mathbf{V}} \equiv \sum_{n=0}^{\infty} \frac{(-t)^n}{n!}
\left(\frac{\partial}{\partial t} + \Lie{\mathbf{V}}{} \right)^n\,.
\]

It can be shown that if $\mathbf{T}$ is any tensor, then
 \[\left(\frac{\partial}{\partial t} +
\Lie{\mathbf{V}}{} \right)(\Proj{\mathbf{V}}\mathbf{T}) = 0 \,,\] that is,
$\Proj{\mathbf{V}}\mathbf{T}$ is the required solution to the \Master  \hspace{0truemm}
for $\mathbf{V}$. In this way, starting from any tensor, we can get Symmetries,
Lagrangians, and other structures, if the value of the projected tensor converges. We
comment finally that if two tensors solve the \Master, then any contraction of their
indices solves it also.

\section{Lagrangian approach to the Heat Equation}
\label{secLagr}
 The Heat Equation (HE), for the field $\omega(x,t)$ and the independent coordinate $x \in
 \mathbb{R}$:

 \begin{equation}\label{heat}
  \omega_t(x,t) = \omega_{xx}(x,t)\,,
  \end{equation}
where $\omega_t \equiv \frac{\partial \omega}{\partial t}$ and $\omega_{xx} \equiv
\frac{\partial^2 \omega}{\partial x^2}$, is represented by the vector
 \[V^{\omega(x)} \doteq \omega_{xx}(x,t)\,.\]
From here on, we drop the label $\omega$ for all the tensors in the $\omega$-coordinate
system only when there is no ambiguity.
 In this way, the above equation reads:
 \[V^x \doteq \omega_{xx}(x,t)\,.\]
 We will project a LB for $V$ starting
from the 2-form ``anti-derivative operator"\cite{derivOp} (which is clearly not a LB
for $V$):
 \[\sigma \opeq D^{-1}\,.\]
As $\sigma$ is not explicitly time-dependent, the projection $\Proj{V}\sigma$ involves
only multiple Lie derivatives along $V$.
 It can be shown that, for $n \geq 0$:
 \[(\Lie{V}{})^n{\sigma} \opeq 2^n D^{2n-1}\,.\]
We sum to get the projection, finding:
 \[\Sigma^{\SS [1]} \equiv \Proj{V}\sigma \opeq
\sum_{n=0}^{\infty} \frac{(-2t)^n}{n!}D^{2n-1}\,,\]
 or
 \[ \Sigma^{\SS [1]} \opeq \eexp{-2t D^2} D^{-1}\,.\]
The inverse\footnote{We define the (up-to-the-kernel) inverse of a $(0,2)$ tensor
$\Sigma$ as a $(2,0)$ tensor $J$ in terms of a contraction: $J^{x z} \Sigma_{z y}
\doteq \delta(x-y)$.} of this operator is a time-dependent PB (which, unfortunately,
does not define a Hamiltonian Structure \cite{hojman2}):
 \[J_{\SS [1]} \equiv \left(\Sigma^{\SS
[1]}\right)^{-1} \opeq D \eexp{2t D^2}.\] Notice that this PB has a finite dimensional
Kernel, generated by the 1-form with constant components.

We build a Lagrangian for the HE contracting the vector $V$  with $\Sigma^{\SS [1]}$,
obtaining
\[
\begin{array}{ccl}
 L^{\SS [2]}_x &\equiv & \Sigma^{\SS [1]}_{x y} V^y \doteq \eexp{-2t D^2} \omega_x(x,t)\\
 \delta L^{\SS [2]} & \opeq & -2 D \eexp{-2t D^2}\,.
 \end{array}
\]
 This defines a new LB:
 \[ \Sigma^{\SS [2]} \equiv \delta L^{\SS [2]} \opeq  -2 D \eexp{-2t D^2} \,.\]
 Notice that this new LB has a finite dimensional Kernel, generated by the constant Symmetry, while
$\Sigma^{\SS [1]}$ does not have a Kernel, and thus the first LB defines a good Action
Principle for the HE, with Lagrangian 1-form given by:

\begin{equation}\label{newLagr}
\begin{array}{rcl}
  L_x^{\SS [1]}& \doteq &- \frac{1}{2} \eexp{-2t D^2} D^{-1}\omega(x,t)\\
  \delta L^{\SS [1]}& = &\Sigma^{\SS [1]}\,,
\end{array}
\end{equation}
and the Action is:

\begin{equation}\label{newAction}
 \mathcal{S} =\frac{1}{2}
\int_{\SS t_0}^{\SS t_1}dt~\int dx~\left(\eexp{-2t D^2} D^{-1} \omega(x,t)\right)\big(
\omega_t(x,t) - \omega_{xx}(x,t) \big)\,.
\end{equation}
The limits in the $x$-integration are $[-a,a]$, with $0<a<\infty$ (HE in the circle) or
$a = \infty$ (HE in the real line).

It is easy to see that these two LB ($\Sigma^{\SS [1]}$ and $\Sigma^{\SS [2]}$) for $V$
are part of a hierarchy of LBs for the HE, for consider the Strong Symmetry\footnote{A
Strong Symmetry is a $(1,1)$ tensor solution of the \Master.} for $V$: $\Lambda_{\SS
[1]} \opeq D.$

This is a known Strong Symmetry for $V$ (see \cite{mauricio}). If we contract it twice
with our (anti-symmetric) $\Sigma^{\SS [1]}$ we obtain, apart from a  constant factor,
$\Sigma^{\SS [2]}$, which is also anti-symmetric and closed. Every two contractions
with $\Lambda_{\SS [1]}$ or $(\Lambda_{\SS [1]})^{-1}$ will give us an anti-symmetric
$(0,2)$ closed tensor, solution of the \Master; in other words, we get a hierarchy or
succession of LBs (and therefore Action Principles) for the HE.

\section{The Metric Structure for the Heat Equation}
\label{secMS}
 Consider now just one contraction to obtain a \emph{symmetric} $(0,2)$ tensor,
solution of the \Master:
\[
  G^{\SS [1]}\equiv \Sigma^{\SS [1]} \cdot \Lambda_{\SS [1]} \opeq \eexp{-2t D^2}\,,
\]
which has no Kernel. Its inverse is the $(2,0)$ tensor:
\[
  G_{\SS [1]} \opeq \eexp{2t D^2}\,.
\]
We define a Metric Structure (MS) as a pair of  solutions of the \Master
\hspace{0truemm} for $V$: a --hopefully invertible-- symmetric $(2,0)$ tensor $G$
(Metric) and a Constant of Motion zero-form $M$ (Metric- or M-Hamiltonian), such that:

\begin{equation}\label{metricStructure}
    V^{x}  =  G^{x y} \delta M_{y}\,,\
\end{equation}
where $ \delta M_{y} \equiv M_{,y} \doteq \frac{\delta M[\omega]}{\delta
\omega(y,t)}\,. $

Equation (\ref{metricStructure}) is analogous to that which appears in the definition
of a Hamiltonian Structure \cite{mauricio, hojman1, hojman2}, but we see that in a MS
the dynamics of any zero-form $C$ is now:
 \[\frac{{\rm d} C}{{\rm d} t} =
\frac{\partial C}{\partial t} + \{C,M\}\,,\]
 where $\{C,M\} \equiv C_{, x}G^{x y}
M_{,y}$ is the anti-commutator. We remark that the M-Hamiltonian $M$ itself must be
an \emph{explicitly} time-dependent Constant of Motion in general, since 
 $ \frac{{\rm d} M}{{\rm d} t} = 0$ and thus $ \frac{\partial M}{\partial t} = \frac{{\rm d} M}{{\rm d} t} - \{M, M\} = - \{M, M\} \neq 0\,$.

 For the Metric \,$G_{\SS [1]}$, the M-Hamiltonian is shown to be:

\begin{equation}\label{hamil}
    M^{\SS [1]}[\omega]  \doteq  -\frac{1}{2}\int~dx~\left(\eexp{- t D^2}\omega_x(x,t)\right)^2 \,. \
\end{equation}

 \section{The General Solution to the Heat Equation and related tensors}
 \label{secGeSo}
If we look at equation (\ref{hamil}), the definition of the M-Hamiltonian $M^{\SS [1]}$
is given in terms of $\eexp{- t D^2}$, an operator with $x$--derivatives of infinite
order, acting  on  the  field  $\omega(x,t)$.  In  order  to  deal  with  these
infinite order derivatives, we are tempted to change our coordinates to
\[
  \tau(x,t) \equiv \eexp{- t D^2}\omega(x,t)\,.
\]
With this change of coordinates, the HE becomes for $\tau$:

\begin{equation}\label{newHeat}
\frac{{\rm d} \tau}{{\rm d} t}(x,t) = 0\,,
\end{equation}
with the general solution \[\tau(x,t) = \tau^{\SS [0]}(x)\,.\] We note that the
M-Hamiltonian of the last section is written now:
 \[ M^{\SS [1]}[\tau]  \doteq  -\frac{1}{2}\int~dx~\big(\tau_x(x,t)\big)^2 \,, \]
which is naturally a constant of the motion for the dynamics (\ref{newHeat}).

Finally, the general solution of the Heat Equation is: \[\omega^{\SS [0]}(x,t) =
\eexp{t D^2}\tau^{\SS [0]}(x)\,.\] All this may seem obvious, but it is very powerful.
For example, if $\tau^{\SS [0]}(x)$ is a polynomial in the variable $x$, we get
immediately polynomial solutions for the HE. It works even with distributions: take
$\tau^{\SS [0]}(x)=\delta(x-s)$, and the associated solution to the HE is the known
solution $\omega^{\SS [0]}(x,t) = \frac{\eexp{-\frac{(x-s)^2}{4t}}}{\sqrt{4 \pi t}}$.

In the next subsections, structures for the HE are presented.

\subsection{Symmetries for the Heat Equation}
 First, we note that any Symmetry for the HE is also a solution of
it (because it is a linear equation), thus we may say that \[\eta^{x} \doteq \eexp{t
D^2} F[\tau](x)\,,\] with $F$ an arbitrary functional of $\tau$ and a function of $x$
only (not of $t$), is a Symmetry for $V$.

As an example, let us suppose that $F[\tau](x) = \int dy\, f(x,y)\tau(y)$, where
$f(x,y)$ is an arbitrary function of its arguments. Then, assuming periodic boundary
conditions or vanishing limiting values for the fields $\omega(x,t)$ and $f(z,x)$ if
$x$ is in the circle or the real line, respectively, we find that:

\begin{equation}\label{symHeat}
 \eta^x \doteq \int dy\,N(x,y,t) \omega(y,t)
\end{equation}
is a linear, nonlocal, Symmetry for the Heat Equation provided
 \[\left(\partial_t - \partial_{xx} +
\partial_{yy}\right)N(x,y,t) = 0\,,\]
and $N(x,y,t)$ is periodic in $y$ or has vanishing $y\to \pm \infty$ values if $y$ is
in the circle or the real line, respectively.

We will use these Symmetries, choosing $\partial_t N = 0$, for the construction of the
Hamiltonian Structure in section \ref{secPB}.

We can consider equation (\ref{symHeat}) as a solution of the HE, so this becomes a way
to find new solutions for the HE starting from known ones.

Observe that the known local Symmetries $\eta_{(l)}^x \doteq \frac{\partial^l}{\partial
x^l} \omega(x,t)$ of the Heat Equation are just special cases of the above
construction, with $N(x,y,t) = \delta^{(l)}(x-y)$.

\subsection{Lagrangians for the Heat Equation}
\label{subsecLagr}
 Now, recalling the definition of a Lagrangian 1-form (\ref{lagrangian}), when we
apply it to the HE it reads:
 \[\bar{\dot{L}}_{x}+ \partial_{xx} L_{x} = 0\,, \]
where the barred dot indicates a total time derivative taken \emph{on-shell\,}. If we
set
 \[L_{x} \doteq \eexp{-t D^2}P[\tau](x)\,,\]
 with $P$ an arbitrary functional, then $L$ is a Lagrangian for $V$.
Thus we see that the Lagrangians constructed in section \ref{secLagr} are obtained by
setting $P^{\SS [1]}[\tau](x) = \tau_x(x)\,$, $\,P^{\SS [2]}[\tau](x) = -\frac{1}{2}
\tau(x)$ .

As an example, we show a construction similar to the one given for the Symmetry. Let us
set $P[\tau](x) = \int dy\,p(x,y) \tau(y)$, so that the Lagrangian reads:

\begin{equation}\label{lagrHeat}
 L_x \doteq \int dy\,Q(x,y,t) \,\omega(y,t)\,,
\end{equation}
where $Q$ is any solution of the $(2+1)$-dimensional negative-time Heat Equation:
 \[\left(\partial_t + \partial_{xx} +
\partial_{yy}\right)Q(x,y,t) = 0\,,\]
and we impose boundary conditions on $Q(x,y,t)$ for the $y$ variable just as those for
$N(x,y,t)$ in the example for Symmetries. The LB which comes from the Lagrangian
(\ref{lagrHeat}) is just
 \[\Sigma_{x y} \doteq Q(y,x,t) - Q(x,y,t)\,, \]
and we have then Action Principles for the HE modulo the Kernel of the  tensor above.

\subsection{Lagrange Brackets and Metrics for the Heat Equation}\label{subsecMetrics}

The $(0,2)$ tensor defined by the operator:
\[
  R \opeq \eexp{-t D^2} \tilde{R}[\tau] \eexp{-t D^2}\,,
\]
where $\tilde{R}[\tau]$ is \emph{any} integro-differential operator depending on
$\omega$ and time only through $\tau$, is a solution of the \Master  \hspace{0truemm}
for the Heat Equation. The proof is straightforward. We get a LB for the HE ($R$
anti-symmetric and closed) if $\tilde{R}[\tau]$ is closed in the $\tau$-coordinate
system. We get a MS for the HE if $\tilde{R}[\tau]$ is a symmetric tensor, such that

\begin{equation}\label{HamStruc}
\partial_{yy}  \tilde{R}[\tau]_{\tau(x) \tau(y)} + \int dz\,\tau_{zz} \tilde{R}[\tau]_{\tau(x)
\tau(z)\,,\,\tau(y)}-\left( x \leftrightarrow y \right) = 0\,,
\end{equation}
so that an M-Hamiltonian can be defined in\footnote{We are assuming we may write $R^{x
z} R_{z y} \doteq \delta(x-y)$, otherwise the equations would get extra terms, related
to the Kernel of the operators $R^{x y}$ and $R_{x y}$.}
\[
 \begin{array}{ccl}
   V^{x} & = & R^{x y} \delta M_y\,.\
 \end{array}
\]
For example, a Metric Structure is defined when $\tilde{R}[\tau]_{\tau(x) \tau(y)}
\doteq \tilde{R}^{\rm o}(x,y)$ is a symmetric and $\tau$-independent solution of:
\[
\left( \partial_{xx} - \partial_{yy}\right)\tilde{R}^{\rm o}(x,y) = 0\,,
\]
and the M-Hamiltonian is
\[
M^{\rm o}[\omega] = \frac{1}{2} \int dx\int dy\,\tilde{R}^{\rm o}(x,y)\left(\eexp{-t
D^2} \omega(x,t)\right) \left(\eexp{-t D^2} \omega_{yy}(y,t)\right)\,.
\]
The MS found in section \ref{secMS} is a particular case of the above construction,
with $\tilde{R}^{\rm o}(x,y)=\delta(x-y)$, the Dirac Delta distribution.

\subsection{Poisson Brackets and Hamiltonian Structures for the Heat Equation}
\label{secPB}
  It should be clear that the $(2,0)$ tensor written as the operator
\[
  J \opeq \eexp{t D^2} \tilde{J}[\tau] \eexp{t D^2}\,
\]
is a Poisson Bracket for the HE provided the $(2,0)$ tensor $\tilde{J}[\tau]$ is
anti-symmetric, satisfies the Jacobi Identity \cite{hojman1, hojman2} in the
$\tau$-coordinate system, and depends on time and $\omega$ only through $\tau$.
Nevertheless, a Hamiltonian Structure is given only when the \emph{partial} time
derivative of $J$ (under constant $\omega$) is zero \cite{hojman2}, in which case the
Hamiltonian H is a time-independent Constant of the Motion and obeys:
 \[ V^{x}  =  J^{x y} \delta H_{y}\,. \]
We consider now an improvement of a time-independent PB
 for the periodic HE (i.e., periodic boundary conditions on the
field $\omega(x,t)$ at $x=-a, \,a$) recently found \cite{mauricio}.
 We use the technique illustrated  in \cite{mauricio, hojman2}.
It starts with a time-independent Constant of the Motion $H^{\rm o}$ (the Hamiltonian)
for the HE, a time-independent Symmetry $\eta_{\rm o}$ such that $\Lie{\eta_{\rm
o}}{H^{\rm o}} \neq 0$ and it needs a set $\ti$ of time-independent Symmetries such
that:
 \[\begin{array}{rcl}
   \Lie{\eta}{H^{\rm o}} &=& 0 \quad \forall \quad \eta \in \ti\,,\\
   \Lie{\eta}{\eta_{\rm o}}&=&0\quad \forall \quad \eta \in \ti\,,\\
   \Lie{\eta}{\bar{\eta}} & = & 0 \quad \forall \quad \eta , \bar{\eta} \in \ti\,.\
 \end{array}\]
After we find these ingredients, we form a Poisson Bracket which, in components, reads:
\[
J^{x y} = \frac{V^x \eta_{\rm o}^y - \eta_{\rm o}^x V^y}{\Lie{\eta_{\rm o}}{H^{\rm o}}}
+ \underset{\eta,\,\bar{\eta} \in \ti}{\sum} \eta^x \bar{\eta}^y - \eta^y
\bar{\eta}^x\,.
\]
 We will take (all integrals in the rest of this subsection are assumed to have limits
$[-a,a]$):
\[
\begin{array}{rcl}
H^{\rm o}& \doteq &\int dx\, \omega(x,t)\\
 \eta_{\rm o}^x~& \doteq& 1\,,
\end{array}
\]
 and look for Symmetries of type (\ref{symHeat}) for the set $\ti$.
If we use time-independent functions (or distributions) $N(x,y)$ such that:

\begin{equation}\label{conditions}
\begin{array}{rcl}
\int dz\,N(z,x)&= &0 \\
 \int dz\,N(x,z)&= &0\\
   \int dz\,\big(N(x,z)\bar{N}(z,y)-\bar{N}(x,z)N(z,y)\big) & = & 0 \\
  N(x,y) & = & A(x+y) + B(x-y)\,,
\end{array}
\end{equation}
with $A$ and $B$ periodic distributions, then $N$, $\bar{N}$ define Symmetries in the
set $\ti$, via equation (\ref{symHeat}).

It is an easy matter to show that the relations

\begin{equation}\label{solutions}
\begin{array}{rcl}
A &=& A_{\SS S}\\
  B & = & B_{\SS S} \\
\int dz\,\big(A(z)+B(z)\big)&=& 0\,, \end{array}
\end{equation}
where $A_{\SS S}(x)\equiv A(-x)$, define solutions of equations (\ref{conditions}) and
thus define Symmetries \linebreak
 $\eta_{\SS A B}^x~\doteq~\int~dy\,~\big(A(x+y)+B(x-y)\big)\omega(y,t)$ in $\ti$. The final
 result, which is the PB for the HE, is:
\[
J^{x y} \doteq \frac{\omega_{xx}(x,t)-\omega_{yy}(y,t)}{2a} + \underset{A ,\,
B,\,\bar{A},\,\bar{B}}{\sum} C_{A B \bar{A} \bar{B}} \int dr\int ds\,\omega(r,t)\,
  \omega(s,t)\, W_{A B \bar{A} \bar{B}}(r,s;x,y)\,,
\]
where $C_{A B \bar{A} \bar{B}}$ are constants,
 \[  W_{A B \bar{A} \bar{B}}(r,s;x,y) = \big[A(x+r)+B(x-r)\big]
\big[\bar{A}(y+s)+\bar{B}(y-s)\big] - x \leftrightarrow y\,,\] and the sum runs over
the space of distributions defined by equations (\ref{solutions}), normalized in the
sense that their first non-zero Fourier coefficient is equal to $1$. In this way, we
have got a true Hamiltonian Structure\footnote{It is an improvement of the PB found in
\cite{mauricio}, which is obtained by setting $C_{A B \bar{A} \bar{B}} = 0$ unless $A =
\bar{A} = 0$, $B(\alpha) = \delta^{(2 n)}(\alpha)$, and $\bar{B}(\alpha) = \delta^{(2
\bar{n})}(\alpha)$, where $n, \bar{n} \in \mathbb{N}$ and $\delta^{l}(\alpha)$ is the
$l$-th derivative of the Dirac Delta distribution.} starting from the trivial general
solution of the HE, and we claim it can be made of maximal rank, because the condition
of time-independence for the PB, inherited by the Symmetries in the set $\ti$, makes
$N(x,y)$ be independent of time, and the solutions to equations (\ref{conditions}) seem
to be exhausted by the set (\ref{solutions}), at least in terms of the dimension of the
space of solutions (for there is another type of solution of equations
(\ref{conditions}), namely (\ref{solutions}) but with $A_{\SS S} = - A$). Of course, we
should be able to compute an invertible  $J$ as well as its inverse if our claim was
right.

 \subsection{Strong Symmetries for the Heat Equation}
 Any $(1,1)$ tensor written as the operator
\[
  \Lambda \opeq \eexp{t D^2} \tilde{\Lambda}[\tau] \eexp{-t D^2}\,,
  \]
where the $(1,1)$ tensor $\tilde{\Lambda}[\tau]$ depends on time and $\omega$ only
 through $\tau$, is a Strong Symmetry for the HE.
For example, if we set $\tilde{\Lambda}_{\SS [1]}[\tau] \opeq D$ and
$\tilde{\Lambda}_{\SS [2]}[\tau] \opeq x$ we obtain, respectively:
\[
  \begin{array}{rcl}
    \Lambda_{\SS [1]} & \opeq & D \\
    \Lambda_{\SS  [2]} & \opeq &2 t D + x \,,\
  \end{array}
\]
which are known Strong Symmetries for the HE \cite{mauricio}.

\section{The Potential Burgers Equation}
\label{secPBE}
 If we make the change of coordinates (Cole-Hopf \cite{cole, hopf} transformation):
 \[ \omega(x,t) =
\eexp{u(x,t)}\,,\]
 we map the Heat Equation (\ref{heat}) into the Potential Burgers Equation (PBE):
\[
 u_t(x,t) = u_{xx}(x,t)+[{u_x}(x,t)]^2\,.
\]
The structures just defined for the Heat Equation are mapped in the following way:

\subsection{Symmetries for the Potential Burgers Equation}
  Any Symmetry of the PBE is written as:
\[
\eta^{u(x)} \doteq \eexp{-u(x,t)} \eexp{t D^2} F[\tau](x) \,, \quad F \,\, {\rm
arbitrary}.
\]
Let us map the example for the HE, see equation (\ref{symHeat}), into the PBE. We end
with a Symmetry for the PBE:
\[
\eta^{u(x)} \doteq \int dy\, N(x,y,t)\,\eexp{u(y,t)-u(x,t)}\,,
\]
where $\left(\partial_t - \partial_{xx} +
\partial_{yy}\right)N(x,y,t) = 0\,.$

\subsection{Lagrangians for the Potential Burgers Equation}

  The 1-form defined as:
\[
L_{u(x)} \doteq \eexp{u(x,t)} \eexp{- t D^2} P[\tau](x)\,, \quad P \,\, {\rm
arbitrary},
\]
is a Lagrangian for the PBE.

The example for the HE, equation (\ref{lagrHeat}), is mapped on a Lagrangian for the
PBE:
\[
  L_{u(x)} \doteq \int dy\,Q(x,y,t) \,\eexp{u(y,t)+u(x,t)}\,,
\]
where
 \[\left(\partial_t + \partial_{xx} +
\partial_{yy}\right)Q(x,y,t) = 0\,.\]
Finally, we map the example for the HE, equations (\ref{newLagr}) and
(\ref{newAction}), into the PBE:
\[
  L_{u(x)}^{\SS [1]} \doteq - \frac{1}{2} \eexp{u(x,t)} \eexp{-2t D^2} D^{-1}\eexp{u(x,t)}\,,
\]
and the Action is:
\[
 \mathcal{S} =\frac{1}{2}
\int_{\SS t_0}^{\SS t_1}dt\int dx\,\left(\eexp{u(x,t)}\eexp{-2t D^2} D^{-1}
\eexp{u(x,t)}\right)\big( u_t(x,t) - u_{xx}(x,t)+[{u_x}(x,t)]^{2} \big)\,.
\]
The Euler-Lagrange equations that come from this Action Principle are equivalent to the
PBE.

\subsection{Lagrange Brackets and Metric Structures for the Potential Burgers Equation}

The general Solution of the PBE's \Master  \hspace{0truemm} for a $(0,2)$ tensor is the
operator
\[
  R[u] \opeq \eexp{u} \eexp{-t D^2} \tilde{R}[\tau] \eexp{-t D^2} \eexp{u}\,,
\]
where $\tilde{R}[\tau]$ is a $(0,2)$ tensor defined in the $\tau$ coordinate system. We
have a Lagrange Bracket if $\tilde{R}[\tau]$ is anti-symmetric and closed in the $\tau$
coordinate system.

The analysis for the MS for the PBE is similar to that given in section
\ref{subsecMetrics}: when $\tilde{R}[\tau]$ is a symmetric solution of equation
(\ref{HamStruc}), an M-Hamiltonian $M[u]$ can be defined in:
\[
  \begin{array}{ccl}
    V^{u(x)} & = & R^{u(x) u(y)} \delta M_{u(y)}\,.\
  \end{array}
\]
As it is shown in section \ref{subsecMetrics}, the most simple solution to equation
(\ref{HamStruc}) we get is when $\tilde{R}[\tau]_{\tau(x) \tau(y)} \doteq
\tilde{R}^0(x,y)$ is a symmetric and $\tau$-independent solution of:
\[
\left( \partial_{xx} - \partial_{yy}\right)\tilde{R}^{\rm o}(x,y) = 0\,,
\]
and the M-Hamiltonian is  mapped into the $u$-coordinate system (PBE):
\[
M^{\rm o}[u] = \frac{1}{2} \int dx \int dy\,\tilde{R}^{\rm o}(x,y) \left(\eexp{-t D^2}
\eexp{u(x,t)}\right)\left(\eexp{-tD^2}\eexp{u(y,t)}\left(u_{yy}(y,t)+[u_y(y,t)]^2\right)\right)\,.
\]

\section{Conclusions}

We have obtained, for the Heat Equation (Potential Burgers Equation), the general
solution of the \Master  \hspace{0truemm} for many important tensors, such as
Symmetries, Lagrangians, Poisson and Lagrange Brackets. We think these results, far
from being trivial, as it could seem at first sight for a linear equation, have brought
a better understanding of the structures related to these evolution equations. In
addition to these, the main contributions in this work are: the Projector, an operator
that takes any tensor and gives back a solution of the \Master; the construction of
valid Action Principles for the Heat Equation (Potential Burgers Equation); the
definition of a new structure, namely the Metric Structure, which bears some
resemblance with the Hamiltonian Structure, but is defined in terms of a symmetric
tensor: the dynamics of functions are anti-commutative and are written in terms of a
time-dependent Constant of the Motion (the Metric Hamiltonian); finally, the
construction of a seemingly maximal-rank Hamiltonian Structure for the Heat Equation,
as an improvement of a recently found result.

The above results (e.g., Hamiltonian Structure) may be extended to the Burgers Equation
and others just by mapping the corresponding tensors under the coordinate
transformations. Also, the results for the Heat Equation may be generalized in a direct
way to every linear equation, such as the Schr\"odinger Equation, and to
finite-dimensional linear systems, like the small-oscillations problem with or without
friction and/or gyroscopic term.

\section*{Acknowledgements}

 One of us (M.B.) is deeply grateful to Fundaci\'on Andes (Grant
for Doctoral Studies), for financial support.


\begin{thebibliography}{100}
\bibitem{derivOp} The derivative operator $D$ is represented by the derivative of the Dirac Delta distribution:
$D f(x) = \int dy\,\delta'(x-y) f(y)$ and the anti-derivative operator $D^{-1}$
performs an integration and is represented by the anti-symmetric Heaviside Step
distribution: $D^{-1} f(x) = \int dy\,\epsilon(x-y) f(y)$.
\end{thebibliography}
\end{document}